\documentclass{article}

\usepackage{arxiv}

\usepackage{enumitem}
\usepackage{lineno}
\usepackage{epstopdf}
\usepackage{float} 
\usepackage{multirow}
\usepackage{longtable}
\usepackage{booktabs}
\usepackage{graphicx} 
\usepackage{graphics} 
\usepackage{caption,subcaption}

\usepackage[utf8]{inputenc} 
\usepackage[T1]{fontenc}    
\usepackage{hyperref}       
\usepackage{url}            
\usepackage{booktabs}       
\usepackage{amsfonts}       
\usepackage{nicefrac}       
\usepackage{microtype}      
\usepackage{lipsum}
\usepackage{graphicx}
\usepackage{placeins}
\usepackage{enumitem} 
\graphicspath{ {./images/} }

\title{COVID-19 Imaging Data Privacy by Federated Learning Design: A Theoretical Framework }

\author{
 Anwaar Ulhaq  \\
  Machine Vision and Digital Health Research Group,\\Charles Sturt University, Port Macquarie, NSW, Australia\\
  \texttt{aulhaq@csu.edu.au}
\AND
  Oliver Burmeister\\
Charles Sturt University, Bathurst, NSW, Australia\\
  \texttt{oburmeister@csu.edu.au}  
}

\begin{document}
\maketitle
\begin{abstract}
To address COVID-19 healthcare challenges, we need frequent sharing of health data, knowledge and resources at a global scale. However, in this digital age, data privacy is a big concern that requires the secure embedding of privacy assurance into the design of all technological solutions that use health data. In this paper, we introduce differential privacy by design (dPbD) framework and discuss its embedding into the federated machine learning system.  To limit the scope of our paper, we focus on the problem scenario of COVID-19  imaging data privacy for disease diagnosis by computer vision and deep learning approaches.  We discuss the evaluation of the proposed design of federated machine learning systems and discuss how differential privacy by design (dPbD) framework can enhance data privacy in federated learning systems with scalability and robustness. We argue that scalable differentially private federated learning design is a promising solution for building a secure, private and collaborative machine learning model such as required to combat COVID19 challenge.  
\end{abstract}



\section{Introduction}
COVID-19 pandemic has changed our world and its challenges \cite{ref32}. Today, more than ever, we require a centralised platform and collective approach to facilitate our collaborative research efforts in different scientific disciplines \cite{ref1,ref31}. Artificial intelligence, especially computer vision, has responded strongly to this challenge \cite{ref11}. Various imaging modalities are being processed and analysed for COVID-19 control \cite{ref2,ref3}. These approaches vary from disease diagnosis and prognosis to disease prevention and management based on different imaging modalities like digital chest x-ray radiography (CXR), chest computed tomography (CT) and Lung ultrasound (LUS) \cite{ref4,ref5,ref6}. An ethical study of imaging data requires the privacy, confidentiality and integrity throughout data analysis. However, it seems a tremendous task due to the existing vulnerabilities of traditional machine learning systems that heavily rely on shared datasets for their training. Similarly, new data protection regulation like General Data Protection Regulation (GDPR) \cite{ref7} in the European Union puts restrictions on the move of data outside of their regional territories.  This situation requires a fundamental change in the ways machine learning systems can work collaboratively. 

Privacy by design (PbD) approach \cite{ref8}  ensures that privacy assurance is embedded into a system design lifecycle by default from the beginning to the end. Therefore, any system built without including privacy considerations as a core part of their design process often behaves poor privacy control. A post (GDPR) world requires a focus on privacy by design with data privacy at the core of system design \cite{ref12} . PbD has a special place in the design of machine learning systems, especially deep learning systems that are going to share our digital future. Google introduced a collaborative machine learning system \cite{ref9,ref10,ref18} with embedded PbD named as federated learning in 2016 that nullified the need for a centralised training data. The fundamental idea is to use different clients or nodes to train local machine learning models on local data samples without any exchange, and sharing of model parameters (e.g. the weights and biases) between these local nodes at some frequency to generate a global model by a process called federated averaging.  All clients or nodes then share this global model. As no data sharing takes place between the nodes, federated design simplicity ensures data privacy. These systems are ideally suited for taring machine learning algorithms for COVID-19 control as sensitive COVID-19 health data will not be shared and would remain in the custody of their subjects. Figure 1 illustrates privacy by design approach to machine learning with a  federated model design for COVID-19 detection using CXR. 

\begin{figure}[H]
	\centering
	\includegraphics[width=12 cm]{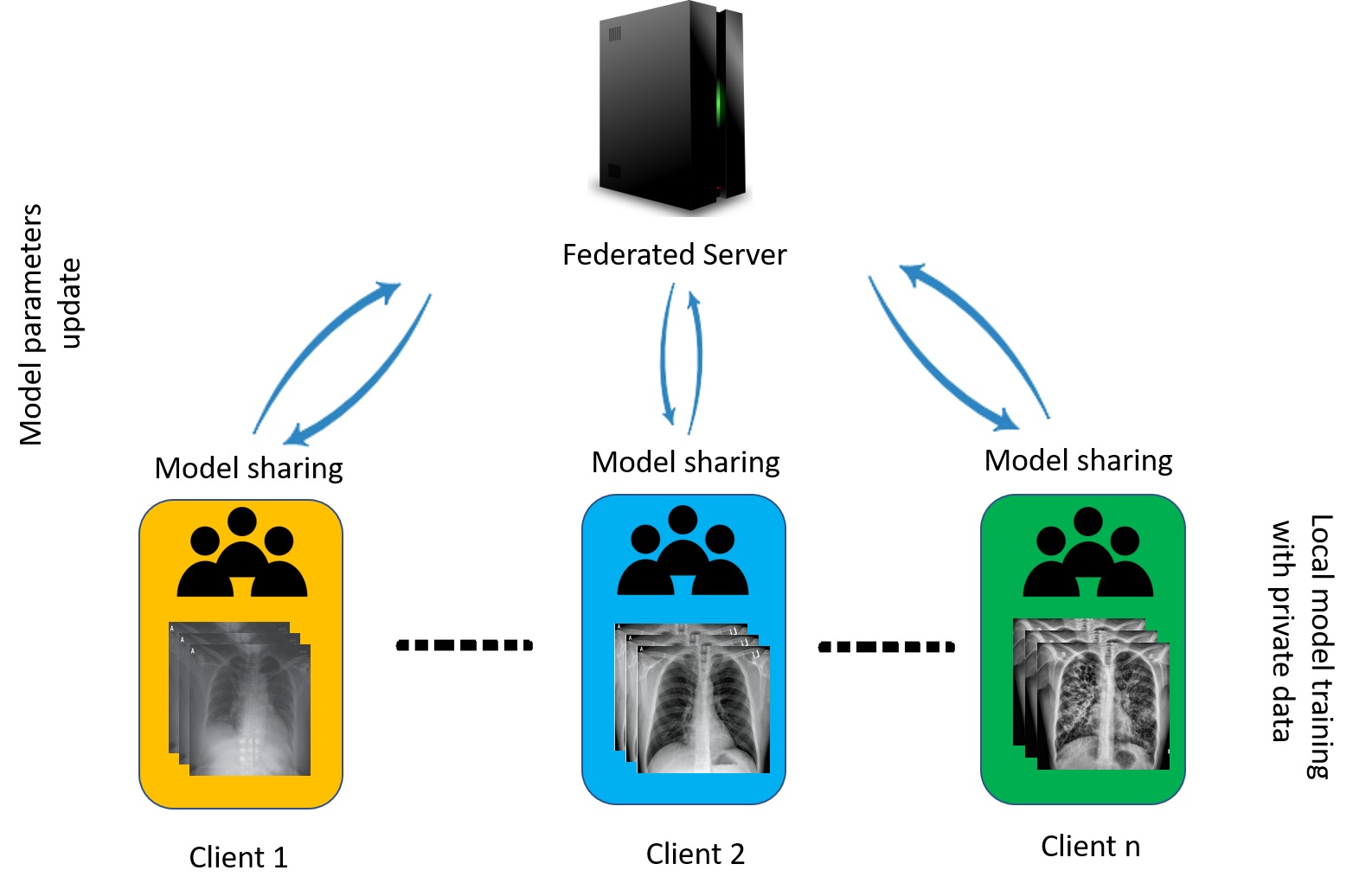}
	\caption{The proposed (dPbD) framework.}
	\label{block}
\end{figure}

However, this process does not guarantee data protection and system robustness. Attackers can figure out individual data based on the global model \cite{ref14,ref17}. Differential privacy adds random noise to an individual’s model, obscuring the results \cite{ref15,ref16}. This random data can be added before the model is shared with the server, without revealing the actual data. This process preserves the individual’s privacy. However, this design approach has limited scalability and robustness. Additionally, this integration is hard to grasp for system designers due to its unreasonable complexity. Inspired from the preliminary empirical works on differentially-private federated learning,  in this paper, we introduce a theoretical framework called differential privacy by design (dPbD) that can help to design scalable and robust federating learning systems for COVID-19data privacy. We address the following research questions in this paper:

\begin{itemize}
\item How can we devise a theoretical framework that underpins all saliant and impeding factors that impact the design of differentially-private federated learning systems with scalability and robustness?
\item How can differentially-private federated learning systems ensure the privacy of COVID-19 imaging data without adversely harming the accuracy?
\end{itemize} 

The contributions of this paper are as follows: 
We extend privacy by design framework in the context of differential privacy and federated learning. We propose a theoretical framework called differential privacy by design (dPbD)  that underpins saliant and impeding factors that impact the design of differentially-private federated learning systems with scalability and robustness. We discuss how the proposed framework can be used to model privacy of COVID-19 imaging data for training differentially-private federated learning systems.

We have organised the paper as follow: Section 2 describes the related work. A brief introduction to preliminary techniques that have inspired the development of this theoretical framework is discussed in section 3. The proposed framework is presented in section 4. We discuss seven principles to implement this design in section 5. Embedding of the proposed principles into federated learning system design for COVID-19 data is discussed in section 6. It is followed by a discussion and concluding remarks.

\section{Related Work}

At an early stage of COVID-19, authorities recognised the importance of privacy and public trust to fight against the pandemic. Some data privacy experts also dubbed it as an unusual scenario like the European Data Protection Board that highlighted article 9 of the General Data Protection Regulation.  This article allows the processing of personal data “for reasons of public interest in the area of public health, such as protecting against serious cross-border threats to health \cite{ref1}. However, careful data-management practices in our data-intensive world require giving critical consideration to data privacy aspects. COVID-19 poses various challenges to the Biomedical Imaging (BI), Artificial Intelligence (AI), Data Analytics (DA), Computer Vision (CV) and Machine learning (ML) as these disciplines require adequate access to big data to detect, diagnose, and predict the spread of the infection \cite{ref2,ref5,ref6}. This article restricts its scope to data privacy of COVID-19 medical imaging data required by computer vision and machine learning approaches for COVID-19 control.

Privacy by design is considered as an approach to embedding privacy directly into system design and introduced by  Ann Cavoukian in 2009 \cite{ref8,ref20}. This design is incorporated into the European GDPR that shows its importance \cite{ref7}. Privacy-preserving machine earning uses privacy by design approach in its framework \cite{ref19}. Most recently, Google introduced a federated machine learning framework for preserving the privacy of the training data \cite{ref9,ref10}. Federated learning systems use distributed framework where model training is performed locally at the client-side, then the updated model parameters are sent to a central server for aggregation. Contray to the idea of data fusion \cite{ref13}, a global model aggregation or fusion is used. Several works \cite{ref17,ref18} discuss design, challenges and future direction in the field of federated learning. We refer the interested readers to these articles. Federated learning systems inherently support privacy by design. 

However, PbD is criticised for being unclear, complex to enforce its implementation, and difficult to adapt to certain disciplines \cite{ref12}.  Anotherline of work in the privacy domain takes a huge interest in leveraging differential privacy into machine learning design for privacy-preserving AI.  We refer review articles \cite{ref15, ref16} for the interested reader. With the increased popularity and enhanced performance of deep learning, difference privacy is proposed for deep learning models and federated learning \cite{ref21,ref22,ref24}. The integration of differential privacy and federated learning leads future research directions in privacy preserving AI. However, this marriage is incredibly complex and lacks clarity in its design due to various competing factors. The complexity of the preliminary work in this direction inspired us to propose a theoretical framework called differential privacy by federated design. It is specially designed to provide differential privacy to COVID-19 imaging data using a federated machine learning system.

\section{Preliminaries:}

\subsection{Federated Machine Learning Systems}

Federated learning is a powerful framework that is devised for machine learning scientists to work collaboratively with decentralised data with privacy by default setting. Google generated the initial idea as part of the series of works in 2015 and 2016 \cite{ref9,ref10}. The initial focus of the federated design was on on-device federated learning tailored to distributed mobile-user interactions. A sample example is G-board app that Predicts the next word to make typing effortless based on typed text using a Federated Ruccerent neural network (RNN) model \cite{ref26}.

Consider a scenario of n machine learning scientists $\{s_{1}, . . ., s_{n} \}$ working on COVID-19 imaging data, all of whom wish to collaborate to train a machine-learning model by sharing their respective imaging data  $\{d_{1}, . . ., d_{n} \}$.  A federated-learning system defines a distributed machine learning process in which the ML scientists collaboratively train a shared model Mf. However, this process ensures that any ML scientist $s_{i}$ does not expose its data $d_{i}$ to any other scientist $s_{j}$ during the local training process. Each of the data scientists, however, shares its local model $m_{i}$ to the server (centralised architecture) \cite{ref23} or on a blockchain ( decentralised architecture) \cite{ref24}. A federating averaging is used for model aggregation to get a  shared model Mf that is then sent back to each ML scientist.

Various improvements are proposed in the literature after the initial design \cite{ref9,ref10}. It includes horizontal and vertical federated learning architectures, federated transfer learning, federated domain adaptation, federated adversarial learning, improving communication protocols and security and making federated learning more personalisable. It is emerging as a promising research topic in machine learning. Interested readers, please refer to the topic reviews \cite{ref17,ref18}.

\subsection{Differential Privacy in Machine Learning}

Differential privacy (DF) defines a formal assurance of anonymity and indistinguishability in terms of a privacy budget ($\epsilon$)—the smaller the budget, the stronger the confidence on privacy \cite{ref15,ref16}. The topic has its roots in aggregate or adjacent databases. In the case of COVID-19 imaging datasets, each COVID-19 imaging training dataset is a set of image-label pairs for supervised learning. Any two of these datasets are adjacent if they vary in a single entry if one image-label pair is present in one image dataset and absent in the other. In other words, Two databases $X$ and $Y$ are neighbours or adjacent if $A(X, Y)=1$   where A is the distance. A randomized mechanism $K:D\rightarrow R$ with
domain D and range R preserves ($\epsilon, \delta$ differential privacy,  if for any pair of adjacent databases $(X, Y)$  belonging to $D$  and set S of possible outputs:$Pr[K(X) \in S] \leq \epsilon Pr[K(Y) \in S] + \delta$.

A randomised mechanism like Gaussian mechanism(GM) \cite{ref27}approximates a real-valued function $F$ for these neighbouring datasets, and differential privacy can be enforced by adding noise to the model to the scale depending on the sensitivity of this function function $F$. The global sensitivity $GS$ of a function $F$ is defined as: $GS_{F}=max_{X,Y:D(X,Y)=1} |F(X)-F(Y)|$.

DF provides a mathematically provable guarantee of privacy protection against a wide range of privacy attacks (include differencing attack, linkage attacks, and reconstruction attacks), Decreasing in epsilon leads to a decrease inaccuracy. $\epsilon$  is a metric of privacy loss at a differentially change in data (adding, removing one entry). The smaller the value is, the better privacy protection while accuracy is defined to be the closeness of the output of DP algorithms to pure production. $F(X)$ can be released accurately when F is insensitive to individual models \cite{ref28}. Many natural functions have low GS like sample mean and covariance matrix. To achieve a small global sensitivity, the ideal condition is that all the clients use sufficient local datasets for training \cite{ref22}.


\section{The proposed Differential Privacy by Design (dPbD) framework  }

In privacy-preserving machine learning, we search for an algorithm that takes as input a dataset (sampled from some distribution), and then privately output a hypothesis h that with high probability has low error over the distribution.

Our proposed theoretical framework is extending privacy by design framework \cite{ref8} for federating machine learning design by using the notion of embedding differential privacy into its system design from end to end. Figure 2 provides a visual description of our proposed differential privacy by design (dPbD) framework.

This framework is partially inspired by empirical studies \cite{ref21,ref22, ref27, ref24, ref23} on the use of differential privacy for machine learning. From these empirical studies, two major dimensions consistently emerged. One dimension differentiates between privacy and utility and highlights their trade-off. Utility in machine learning can be taken as model accuracy or decrease in testing loss—the other dimensions design scalability and robustness. Here scalability means scalability of federated learning model and number of clients. Robustness indicates system performance against attacks. 

There are four quadrants and thus four important junctions of the proposed framework that define most important factors in defining differential privacy in federated learning design. These factors include level of privacy, level of randomised noise, global sensitivity and number of clients or nodes in a differentially private federated learning system. 

As differential privacy guarantees anonymity and indistinguishability in terms of a privacy budget (epsilon)—the smaller the budget, the stronger the confidence on privacy, and we call it a level of privacy. It is significant characteristics of differential privacy by design as it guarantees and provides a quantitative notion of privacy compared to the unclear concept of privacy in privacy by design framework. The level of privacy defines system robustness against attacks. Similarly adding more noise as part of differential privacy can reduce system utility and robustness. The amount of randomised noise is thus an important factor to balance the trade-off between privacy and utility. 

On the other hand, scalability requires an increase in the number of nodes and the effectiveness of robust aggregation. Robust aggregation depends on global sensitivity. A small value of global sensitivity requires that all the clients use sufficient local datasets for training, .and the type of aggregation function. Similarly, functions having low global sensitivity are preferred for better performance and utility like Sample mean and Covariance matrix.

\section{The 7 Foundational Principles }

We advise these seven principles to accomplish proposed differential privacy by design framework for system design. 

\begin{itemize}

\item Privacy can be guaranteed in design: The system must ensure a meaningful privacy guarantee. For instance, choosing a smaller epsilon produces noisier results and better privacy guarantees in differential privacy. Privacy guarantee during system design will build trust. 

\item Privacy can be quantified in design: Differential privacy can be used to quantify privacy. The strategy of using budgets, expenses and losses in terms of privacy is known as privacy accounting. The maximum privacy loss is called the privacy budget. This quantification leads to better privacy-preserving design.

\item Privacy by Modularity: Modularity is the Key to ensure privacy as it reduces complexity \cite{ref29}. Modules can be removed, replaced, or upgraded without affecting other components. Privacy of the system should not be affected by removing, replacing, or upgrading any system component.

\item  Privacy Embedded into Federated Design : Privacy is preserved in model aggregation with low global sensitivity. The aggregated model should also be insensitive to local data at different nodes.

\item Privacy with Scalability :Privacy notion is unaffected by scaling the system. Any change in scalability for Federated Learning system and asynchronous or synchronous training algorithms should not degrade privacy. 

\item Anonymity in Design Lifecycle:	Original data is never shared. Only modified function or model parameters are transmitted. Data anonymity must be ensured in complete Design Lifecycle.

\item Optimising Privacy-Utility Trade-off in Design:	 we refer utility  to certain system properties \cite{ref30}. Increase in privacy often causes a decrease in utility. A design should optimise this trade-off by ensuring privacy without adversely harming system utility.
\end{itemize}

\begin{figure}[H]
	\centering
	\includegraphics[width=16 cm]{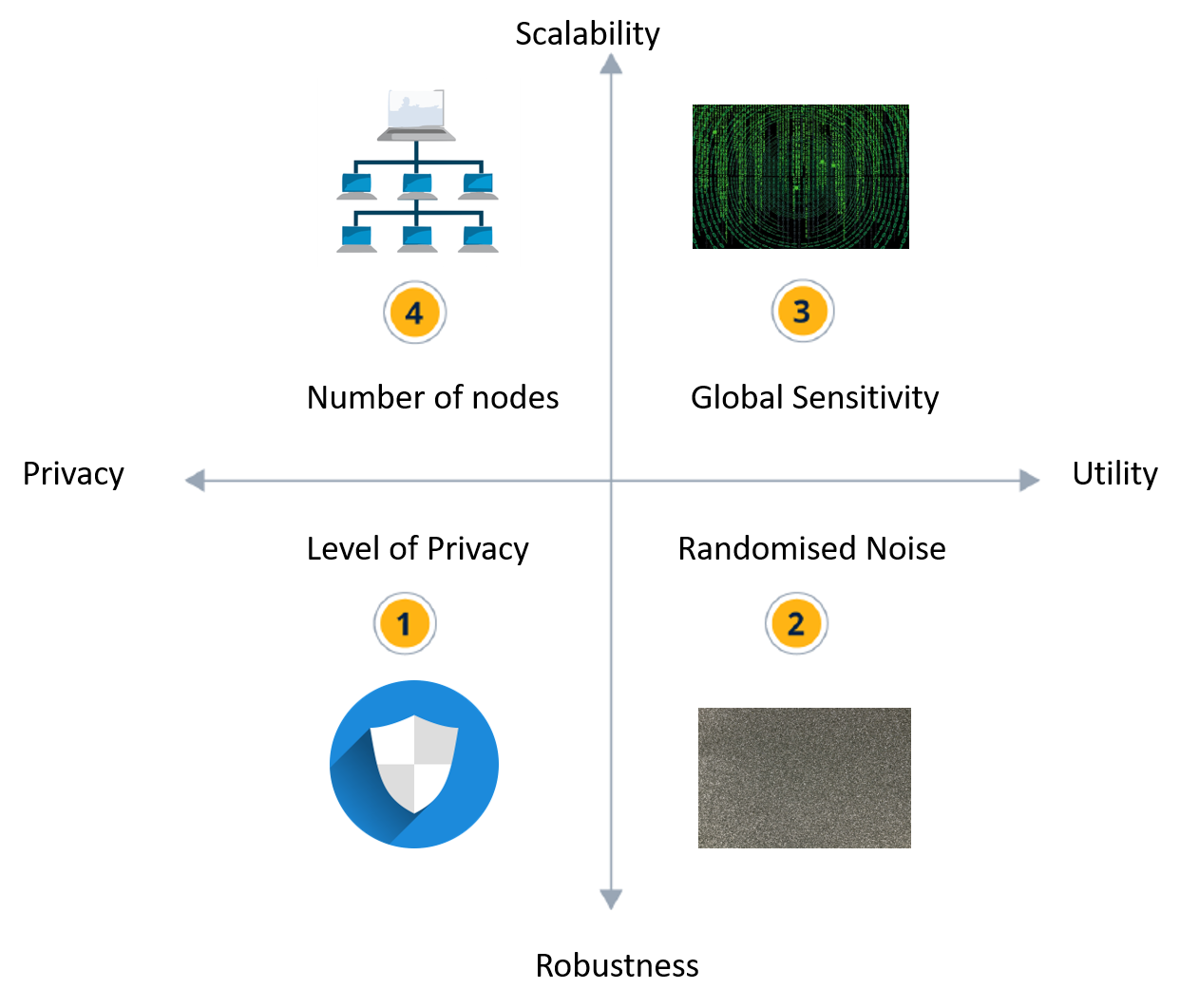}
	\caption{The proposed differential privacy by design (dPbD) framework.}
	\label{block}
\end{figure}

\section{A Scenario of COVID-19 Imagaing Data Privacy By Design}

The privacy of COVID-19  imaging data on a collaborative machine learning network (federated  machine learning) can be taken as a case study of differential privacy by design framework. Its seven foundation principles can be applied to ensure privacy while designing collaborative machine learning systems.  Here, we describe a brief view of the implementation of these seven principles in the scenario of the training machine learning model on federated learning systems.

\begin{itemize}

\item Privacy-preserving federated learning design must control the level of privacy to control perturbation and noise for robustness. However, the design must provide equate level of privacy to prevent patients personal information embedding or related to their COVID-19 imaging data. 

\item Privacy level must be quantified in federated  design. There must be understanding and consensus between teams involved in federated learning setup about the minum and maximum level of privacy required for collaborative research.  

\item Adequate privacy must be enforced during design at all system components like clients, federated server and communication channels. Addition or removal of clients and change in client side learning model should not affect data privacy promised by the system. 

\item A robust model aggregation strategy must be adopted. A function with low global sensitivity be chosen while designing any federated system for COVID-19. The robust aggregation should be insensitive to local changes and changes in data on client nodes. 

\item COVID-19 imaging data is never shared throughout the training lifecycle. Model sharing is protected and maintain the anonymity of patients information through the model up-gradation and communication rounds. 

\item The system design must be flexible, and adding and removing clients and teams or increasing or decreasing COVID-19 data size should not adversely harm patient data and shared model privacy. Horizontal and vertical scalability be supported during design with features of robustness and resilience. 

\item 	Accuracy is an important consideration in COVID diagnosis, and it should not be harmed. It is critical to reduce or sacrifice it for the sake of privacy as it can adversely affect system utility. An optimised trade-off be sought during the design. 

\end{itemize} 

\section{Discussion and Concluding Remarks}

The proposed differential privacy by design framework is focused on the design of privacy-preserving federated machine learnng systems. This theoretical framework is developed by inspiring from various empirical studies about the use of differential privacy in federated learning. However, we discovered that while the majority of the proposed systems emphasise on the trade-off between privacy and utility, they often ignore scalability and robustness of the system. Our proposed framework fills up that gap and proposes a comprehensive framework that covers the majority of the design concepts. We argue that following the footsteps of privacy by design framework, differential privacy must be embedded throughout the design lifecycle, and it should provide overall coverage of protection and privacy to any proposed federated machine learning system. To pave the way of this embedding in design lifecycle, we defined seven foundational principles similar to seven principles of privacy by design framework. 

We use a case study of COVID-19 imaging data. However, our proposed framework should be applicable to all types of data on a federated learning system.  In our future work, we would implement and validate these principles for designing a computer vision-based COVID-19 diagnosis system based on pathological imaging data available from various teams around the world.  We hope that the proposed framework with impact designing a  privacy-preserving federated learning system with reduced complexity and sufficient data protection for collaborative research to combat COVID-19 challenge.


\bibliographystyle{unsrt}  
\bibliography{ABC}  

\begin{thebibliography}{10}

\bibitem{ref32}
Urs Gasser, Marcello Ienca, James Scheibner, Joanna Sleigh, and Effy Vayena.
\newblock Digital tools against covid-19: taxonomy, ethical challenges, and
  navigation aid.
\newblock {\em The Lancet Digital Health}, 2020.

\bibitem{ref1}
Marcello Ienca and Effy Vayena.
\newblock On the responsible use of digital data to tackle the covid-19
  pandemic.
\newblock {\em Nature medicine}, 26(4):463--464, 2020.

\bibitem{ref31}
Sirina Keesara, Andrea Jonas, and Kevin Schulman.
\newblock Covid-19 and health care’s digital revolution.
\newblock {\em New England Journal of Medicine}, 382(23):e82, 2020.

\bibitem{ref11}
Wim Naud{\'e}.
\newblock Artificial intelligence vs covid-19: limitations, constraints and
  pitfalls.
\newblock {\em Ai \& Society}, page~1, 2020.

\bibitem{ref2}
Anwaar Ulhaq, Asim Khan, Douglas Gomes, and Manoranjan Pau.
\newblock Computer vision for covid-19 control: A survey.
\newblock {\em arXiv preprint arXiv:2004.09420}, 2020.

\bibitem{ref3}
Michael~J Horry, Manoranjan Paul, Anwaar Ulhaq, Biswajeet Pradhan, Manash Saha,
  Nagesh Shukla, et~al.
\newblock X-ray image based covid-19 detection using pre-trained deep learning
  models.
\newblock 2020.

\bibitem{ref4}
Michael~J Horry, Subrata Chakraborty, Manoranjan Paul, Anwaar Ulhaq, Biswajeet
  Pradhan, Manas Saha, and Nagesh Shukla.
\newblock Covid-19 detection through transfer learning using multimodal imaging
  data.
\newblock {\em IEEE Access}, 8:149808--149824, 2020.

\bibitem{ref5}
Anwaar Ulhaq, Jannis Born, Asim Khan, Douglas Gomes, Subrata Chakraborty, and
  Manoranjan Paul.
\newblock Covid-19 control by computer vision approaches: A survey.
\newblock {\em IEEE Access}, 2020.

\bibitem{ref6}
Douglas~PS Gomes, Anwaar Ulhaq, Manoranjan Paul, Michael~J Horry, Subrata
  Chakraborty, Manas Saha, Tanmoy Debnath, and DM~Rahaman.
\newblock Potential features of icu admission in x-ray images of covid-19
  patients.
\newblock {\em arXiv preprint arXiv:2009.12597}, 2020.

\bibitem{ref7}
Paul Voigt and Axel Von~dem Bussche.
\newblock The eu general data protection regulation (gdpr).
\newblock {\em A Practical Guide, 1st Ed., Cham: Springer International
  Publishing}, 2017.

\bibitem{ref8}
Ann Cavoukian et~al.
\newblock Privacy by design: The 7 foundational principles.
\newblock {\em Information and privacy commissioner of Ontario, Canada}, 5,
  2009.

\bibitem{ref12}
Ira~S Rubinstein.
\newblock Regulating privacy by design.
\newblock {\em Berkeley Tech. LJ}, 26:1409, 2011.

\bibitem{ref9}
Jakub Kone{\v{c}}n{\`y}, Brendan McMahan, and Daniel Ramage.
\newblock Federated optimization: Distributed optimization beyond the
  datacenter.
\newblock {\em arXiv preprint arXiv:1511.03575}, 2015.

\bibitem{ref10}
Jakub Kone{\v{c}}n{\`y}, H~Brendan McMahan, Daniel Ramage, and Peter
  Richt{\'a}rik.
\newblock Federated optimization: Distributed machine learning for on-device
  intelligence.
\newblock {\em arXiv preprint arXiv:1610.02527}, 2016.

\bibitem{ref18}
Qiang Yang, Yang Liu, Tianjian Chen, and Yongxin Tong.
\newblock Federated machine learning: Concept and applications.
\newblock {\em ACM Transactions on Intelligent Systems and Technology (TIST)},
  10(2):1--19, 2019.

\bibitem{ref14}
Arjun~Nitin Bhagoji, Supriyo Chakraborty, Prateek Mittal, and Seraphin Calo.
\newblock Analyzing federated learning through an adversarial lens.
\newblock In {\em International Conference on Machine Learning}, pages
  634--643. PMLR, 2019.

\bibitem{ref17}
Tian Li, Anit~Kumar Sahu, Ameet Talwalkar, and Virginia Smith.
\newblock Federated learning: Challenges, methods, and future directions.
\newblock {\em IEEE Signal Processing Magazine}, 37(3):50--60, 2020.

\bibitem{ref15}
Cynthia Dwork.
\newblock Differential privacy: A survey of results.
\newblock In {\em International conference on theory and applications of models
  of computation}, pages 1--19. Springer, 2008.

\bibitem{ref16}
Zhanglong Ji, Zachary~C Lipton, and Charles Elkan.
\newblock Differential privacy and machine learning: a survey and review.
\newblock {\em arXiv preprint arXiv:1412.7584}, 2014.

\bibitem{ref20}
Kirsten Wahlstrom, Anwaar Ul-haq, Oliver Burmeister, et~al.
\newblock Privacy by design.
\newblock {\em Australasian Journal of Information Systems}, 24, 2020.

\bibitem{ref19}
Payman Mohassel and Yupeng Zhang.
\newblock Secureml: A system for scalable privacy-preserving machine learning.
\newblock In {\em 2017 IEEE Symposium on Security and Privacy (SP)}, pages
  19--38. IEEE, 2017.

\bibitem{ref13}
AM~Mirza, Sajid Qamar, et~al.
\newblock An optimized image fusion algorithm for night-time surveillance and
  navigation.
\newblock In {\em Proceedings of the IEEE Symposium on Emerging Technologies,
  2005.}, pages 138--143. IEEE, 2005.

\bibitem{ref21}
Martin Abadi, Andy Chu, Ian Goodfellow, H~Brendan McMahan, Ilya Mironov, Kunal
  Talwar, and Li~Zhang.
\newblock Deep learning with differential privacy.
\newblock In {\em Proceedings of the 2016 ACM SIGSAC Conference on Computer and
  Communications Security}, pages 308--318, 2016.

\bibitem{ref22}
Kang Wei, Jun Li, Ming Ding, Chuan Ma, Howard~H Yang, Farhad Farokhi, Shi Jin,
  Tony~QS Quek, and H~Vincent Poor.
\newblock Federated learning with differential privacy: Algorithms and
  performance analysis.
\newblock {\em IEEE Transactions on Information Forensics and Security}, 2020.

\bibitem{ref24}
Rajesh Kumar, Abdullah~Aman Khan, Sinmin Zhang, WenYong Wang, Yousif Abuidris,
  Waqas Amin, and Jay Kumar.
\newblock Blockchain-federated-learning and deep learning models for covid-19
  detection using ct imaging.
\newblock {\em arXiv preprint arXiv:2007.06537}, 2020.

\bibitem{ref26}
Dianlei Xu, Tong Li, Yong Li, Xiang Su, Sasu Tarkoma, and Pan Hui.
\newblock A survey on edge intelligence.
\newblock {\em arXiv preprint arXiv:2003.12172}, 2020.

\bibitem{ref23}
Boyi Liu, Bingjie Yan, Yize Zhou, Yifan Yang, and Yixian Zhang.
\newblock Experiments of federated learning for covid-19 chest x-ray images.
\newblock {\em arXiv preprint arXiv:2007.05592}, 2020.

\bibitem{ref27}
Robin~C Geyer, Tassilo Klein, and Moin Nabi.
\newblock Differentially private federated learning: A client level
  perspective.
\newblock {\em arXiv preprint arXiv:1712.07557}, 2017.

\bibitem{ref28}
Krishna Pillutla, Sham~M Kakade, and Zaid Harchaoui.
\newblock Robust aggregation for federated learning.
\newblock {\em arXiv preprint arXiv:1912.13445}, 2019.

\bibitem{ref25}
Aleksei Triastcyn and Boi Faltings.
\newblock Federated learning with bayesian differential privacy.
\newblock In {\em 2019 IEEE International Conference on Big Data (Big Data)},
  pages 2587--2596. IEEE, 2019.

\bibitem{ref29}
Joanna~J Bryson and Lynn~Andrea Stein.
\newblock Modularity and design in reactive intelligence.
\newblock In {\em International Joint Conference on Artificial Intelligence},
  volume~17, pages 1115--1120. LAWRENCE ERLBAUM ASSOCIATES LTD, 2001.

\bibitem{ref30}
Ali Makhdoumi and Nadia Fawaz.
\newblock Privacy-utility tradeoff under statistical uncertainty.
\newblock In {\em 2013 51st Annual Allerton Conference on Communication,
  Control, and Computing (Allerton)}, pages 1627--1634. IEEE, 2013.

\end{thebibliography}
\end{document}